%% file: unlock_memory_ai.tex
\documentclass[sigconf]{acmart}

\AtBeginDocument{%
  \providecommand\BibTeX{{%
    \normalfont B\kern-0.5em{\scshape i\kern-0.25em b}\kern-0.8em\TeX}}}

\setcopyright{rightsretained}
\copyrightyear{2024}
\acmYear{2024}
\acmDOI{10.1145/3613905.3650979}

\acmConference[CHI EA '24]{Extended Abstracts of the CHI Conference on Human Factors in Computing Systems}{May 11--16, 2024}{Honolulu, HI, USA}

\acmBooktitle{Extended Abstracts of the CHI Conference on Human Factors in Computing Systems (CHI EA '24), May 11--16, 2024, Honolulu, HI, USA}
\acmISBN{979-8-4007-0331-7/24/05}

\begin{document}

\title{Unlocking Memories with AI: Exploring the Role of AI-Generated Cues in Personal Reminiscing}

\author{Jun Li Jeung}
\email{j.l.jeung@student.tue.nl}
\orcid{0009-0008-0510-3718}
\affiliation{%
  \institution{Eindhoven University of Technology}
  \city{Eindhoven}
  \country{The Netherlands}
}

\author{Janet Yi-Ching Huang}
\email{y.c.huang@tue.nl}
\orcid{0000-0002-8204-4327}
\affiliation{%
  \institution{Eindhoven University of Technology}
  \city{Eindhoven}
  \country{The Netherlands}}


\begin{abstract}
    \label{sec:abstract}
    \input{Body/00-abstract}
\end{abstract}

\ccsdesc[500]{Human-centered computing~Interaction paradigms}

\keywords{LLMs, AI for Personal Reminiscing, AI-Generated Memory Cues}

\begin{teaserfigure}
  \includegraphics[width=\textwidth]{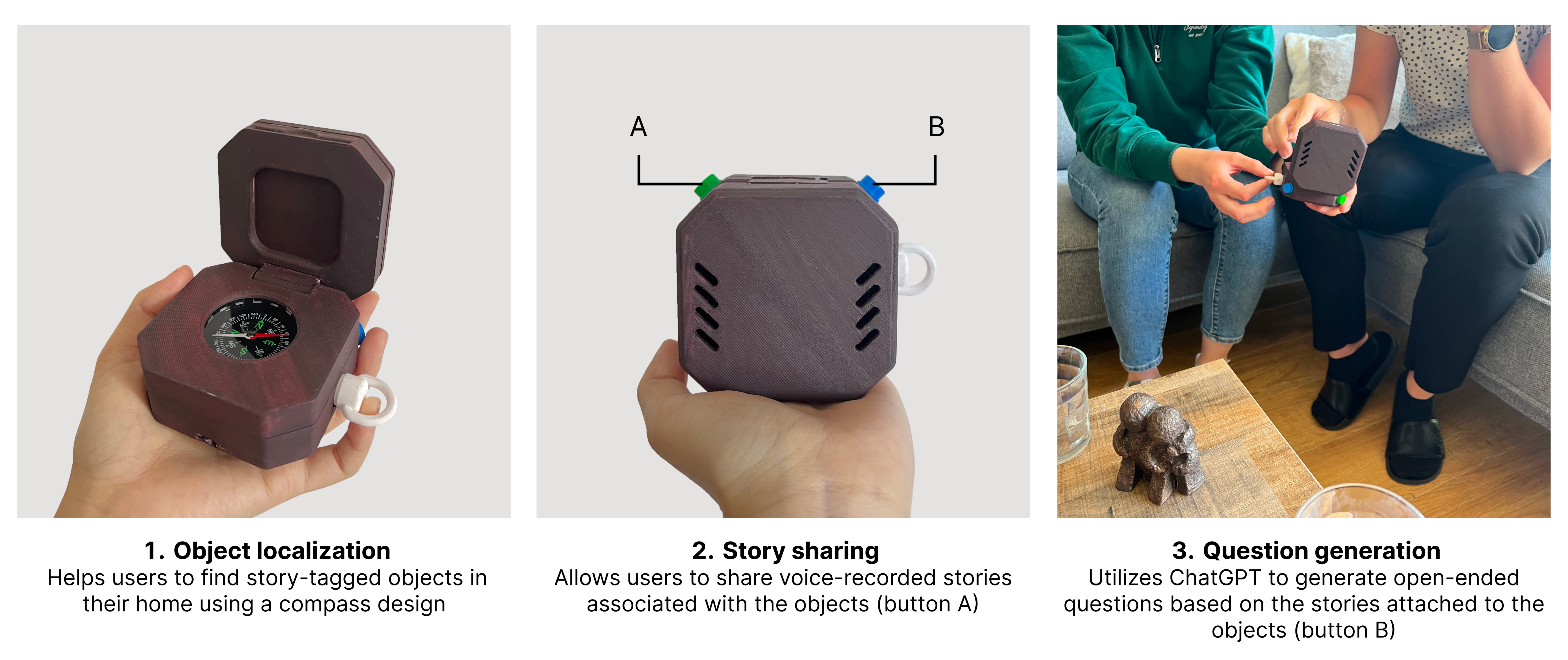}
   \vspace*{-0.8cm}
  \caption{Treasurefinder, step-by-step guide to trigger participants in their remembering processes}
   \vspace*{0.4cm}
  \label{fig:Treasurefinder}
\end{teaserfigure}


\maketitle

\section{Introduction}
\label{sec:intro}
\input{Body/01-introduction}

\section{Related Work}
\label{sec:related-work}
\input{Body/02-related-work}

\section{Treasurefinder}
\label{sec:Treasurefinder}
\input{Body/03-1-Treasurefinder}

\section{Study Design}
\label{sec:Study-study}
\input{Body/03-study-design}

\section{Findings}
\label{sec:Findings}
\input{Body/04-findings}

\section{Discussion and Future Work}
\label{sec:discussion}
\input{Body/05-discussion}

\section{Conclusion}
\label{sec:conclusion}
\input{Body/06-conclusion}

\begin{acks}
This work is supported by the Department of Industrial Design and the Eindhoven Artificial Intelligence Systems Institute at Eindhoven University of Technology.
\end{acks}


\bibliographystyle{ACM-Reference-Format}
\bibliography{unlock_memory_ai}

\end{document}

%% file: Body/00-abstract.tex
While technology-mediated reminiscing has been studied for decades, generating relevant cues to trigger personal reminiscing remains challenging. The potential of AI in generating relevant content across various domains has been recently recognized, yet its use in facilitating reminiscing is still less explored. This work aims to explore the use of AI in supporting the recall of personal memories associated with significant objects at home. We designed Treasurefinder, a device powered by a large language model (LLM) that generates open-ended questions based on stories stored in NFC-tagged physical objects or cards. We conducted an exploratory study with 12 participants, grouped in pairs, to observe reminiscing behaviors when using Treasurefinder. The results showed the AI-generated questions 1) supported individuals to recall the past, 2) provided new insights about the other person, and 3) encouraged reflection. Notably, the device facilitated active memory retrieval related to cherished objects that are often overlooked.

%% file: Body/01-introduction.tex
Reflecting on past experiences is vital in our daily lives. This practice benefits not only individuals with memory issues, like dementia patients but also anyone who wants to engage with their memories. Despite its benefits, reminiscing often goes unnoticed in daily routines, usually happening spontaneously or in response to cues, for example, from tangible objects~\cite{Domenique_2015}. To address this, there has been significant exploration into technology-mediated reminiscing, often involving technologically enhanced artifacts~\cite{frohlich_memory_2000,niemantsverdriet_interactive_2016,kim_slide2remember_2022}. However, a challenge arises as people find it difficult to form attachments to such artifacts. Individuals highly value the physical traces and marks on meaningful objects as tangible reminders of the past, which enhances awareness of the object's historical significance~\cite{zijlema_preserving_2017}. Despite the effectiveness of meaningful objects in triggering reminiscing, integrating technology directly into these objects presents a significant challenge. This work aims to explore a novel way to facilitate reminiscing practices using these meaningful objects.

The development of AI-driven tools, particularly Large Language Models (LLMs), has significantly enhanced data analysis, human-like conversations, and contextual text generation capabilities~\cite{ChatGPT_street_lets_2023}. A study using ChatGPT demonstrated its potential in assisting word retrieval for individuals with language disorders, such as aphasia, which relates to memory retrieval processes~\cite{Kumar_chatgpt_2023}. This evidence underscores the potential of such tools in supporting remembering practices, but this research field lacks extensive exploration. Prior studies have shown that open-ended questions can enhance memory recall~\cite{harris_teaching_2022}. However, crafting relevant open-ended questions is challenging for people, as they often rely on questions based on their personal interests or existing knowledge~\cite{harris_teaching_2022,kulkofsky_why_2010}. Given these challenges, AI, particularly LLMs, could play a crucial role in facilitating memory recall by generating appropriate, contextually relevant questions as cues. As such, this work seeks to explore whether and how integrating AI, especially LLMs, into physically meaningful objects can support people’s reminiscing practices.

In this work, we designed Treasurefinder, a compact device integrated with an LLM that generates thought-provoking, relevant, open-ended questions. When people scan NFC-tagged objects or insert NFC-tagged cards, Treasurefinder generates open-ended questions related to the story stored in the objects, triggering users to recall or reflect on the stories behind the objects. We conducted a field study involving 12 participants, consisting of pairs who knew each other. We observed how participants perceive and interact with Treasurefinder in a home setting.

Our findings indicated that our design encouraged individuals to recall past memories, as these AI-generated questions were both meaningful and relevant. These questions also prompted new insights about others. Furthermore, our design facilitated reflection on past experience. Most interestingly, the device facilitated the active retrieval of memories linked to often overlooked but cherished objects. These findings shed light on the possibilities of using LLMs for reminiscing, as it positively supports memory recollection and promotes reflection.

%% file: Body/02-related-work.tex
\subsection{Reminiscing through artifacts}
Memories are often evoked through specific cues, with diverse stimuli such as images~\cite{kim_slide2remember_2022,li_slots-memento_2019,odom_designing_2014}, sounds~\cite{frohlich_memory_2000,kim_slide2remember_2022,niemantsverdriet_interactive_2016}, texts~\cite{odom_technology_2012}, and tangible objects acting as key triggers for memory recollection. These cues facilitate the retrieval of information from long-term memory. This process has been explored extensively through various technology-mediated artifacts~\cite{frohlich_memory_2000,niemantsverdriet_interactive_2016,kim_slide2remember_2022,li_slots-memento_2019}. For instance, the Memory Box allows users to attach stories to personal objects via voice recordings~\cite{frohlich_memory_2000}. These stories play automatically when the object is removed from the box. Niemantsverdriet~\cite{niemantsverdriet_interactive_2016} has created interactive jewellery that captures ambient sounds to prompt personal reminiscence. Recently, Kim et al.~\cite{kim_slide2remember_2022} designed an interactive wall photo frame that encourages people to engage with digital photos, helping to bring their past experiences to life.

While reminiscing artifacts often incorporate embedded technologies~\cite{frohlich_memory_2000,niemantsverdriet_interactive_2016,kim_slide2remember_2022}, prior research showed that people heavily rely on everyday objects as cues for reminiscence activities~\cite{Domenique_2015}. This finding suggests a need to re-evaluate our design strategy, particularly integrating technology with everyday objects. This strategy should not only respect the innate characteristics of these objects but also preserve their physical traces and marks, which hold sentimental value. Aligning with user preferences ensures the emotional and historical essence of these objects is retained while improving their capability to trigger memories~\cite{zijlema_preserving_2017}.

\subsection{Reminiscing through AI-powered products}
Research on using AI tools for supporting human reminiscing experiences is limited. However, existing studies show the potential of these tools in this area. For instance, a recent preliminary study demonstrated the potential of using ChatGPT~\cite{Kumar_chatgpt_2023} in assisting word retrieval for individuals with language disorders like Aphasia. Nikititina et al. introduce the use of a conversational agent in supporting remembering practices. This agent, leveraging Google Cloud Vision, gathers in-context information from visual images and generates questions related to the images~\cite{nikitina_smart_2018}. Nevertheless, this scripted chatbot has not yet been tested to evaluate its effectiveness in facilitating reminiscence. 

Prior work emphasizes the importance of open-ended questions in enhancing memory recall~\cite{harris_teaching_2022}. However, crafting relevant and engaging questions can be challenging for many people. Most people depend on personal interests or existing knowledge to formulate questions~\cite{harris_teaching_2022,kulkofsky_why_2010}. This is where AI, especially LLMs, can play a pivotal role in processing diverse contextual data and generating human-like responses~\cite{ChatGPT_street_lets_2023}. By generating relevant, context-sensitive questions, LLMs could significantly streamline the process of memory recall, acting as an effective cue in reminiscing practices.

%% file: Body/03-1-Treasurefinder.tex
In this paper, we introduce Treasurefinder, a compact device that incorporates an LLM to generate engaging open-ended questions. The device activates when users scan an NFC-tagged object at home or insert an NFC-tagged card. It then plays voice stories previously recorded by the object's owners, allowing users to interact with the objects and their associated stories. 

Treasurefinder has three key functions: (1)~\textbf{Object localization:} it helps users to find story-tagged objects in their home using a compass design. (2)~\textbf{Story sharing:} it allows users to share voice-recorded stories associated with these objects. (3)~\textbf{Question generation:} it utilizes an LLM to generate open-ended questions based on the stories attached to these objects. Specifically, we used the Chat completions API\footnote{https://platform.openai.com/docs/guides/text-generation/chat-completions-api} from the OpenAI text generation model GPT-3.5 to generate the questions. Next, we converted the LLMs-generated text into speech using the text-to-speech tool eSpeak\footnote{https://espeak.sourceforge.net/index.html}. The LLM-generated questions aim to prompt deeper thoughts and recollections.

To generate reflective questions with an LLM, the following prompt is used: ``First, find relevant information that could spark thoughts about their past. Then, based on the relevant information, ask one question to reflect on the past. Do it from the following text:'' The complete step-by-step instructions are in Figure~\ref{fig:Treasurefinder}.

%% file: Body/03-study-design.tex
We conducted a study involving twelve participants to investigate how incorporating AI into physically meaningful objects can support people’s remembering practices at home. Our design was tested with six pairs of participants who already knew each other. The goal of the study was to explore the role of AI-generated questions in promoting relationship-building and reflection.

\subsection{Participants}
In this study, we recruited 12 participants, ages 23 to 30, without prior memory issues. Of these, 5 were male and 7 were female. We intentionally recruited participants in pairs who already knew each other. These pairs were categorized into three groups: friends (F), couples (C), and family-related (siblings) (FR). For anonymity, we will refer to both pairs and individuals by their assigned IDs. The pairs FA, FB, CA, CB, FRA, and FRB were treatment pairs, where ``-1'' and ``-2'' indicate the individual participant in each pair, such as FA-1 or FA-2. The number `1' and `2' represent the role of homeowner and guest, respectively.

\begin{figure*}
  \centering
  \includegraphics[width=\linewidth]{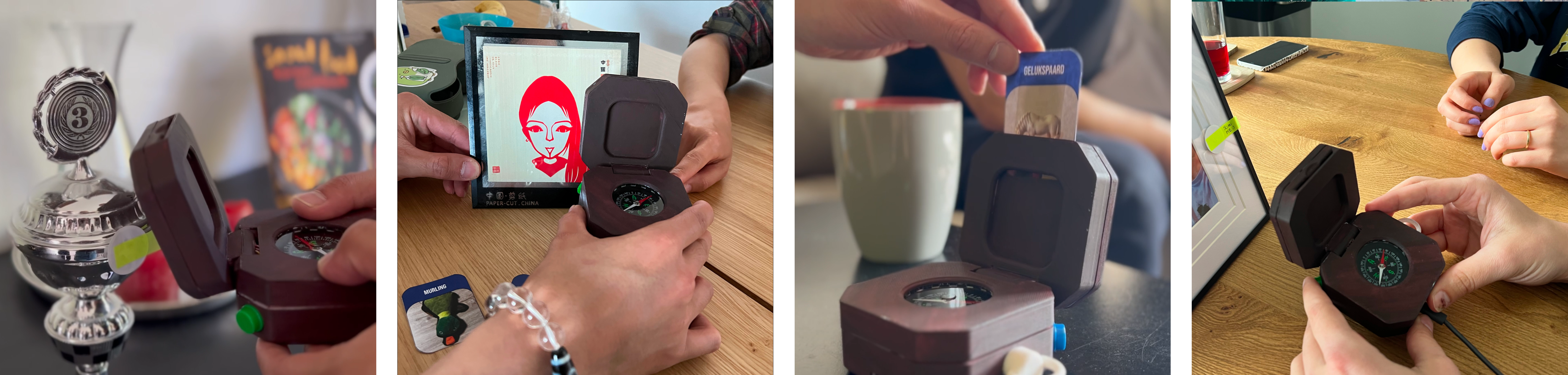}
  \caption{User Testing at diverse homeowners' location. Participants insert cards or scan meaningful objects and then listen to either the voice-recorded stories or the AI-generated open-ended questions.}
  \label{fig:Testing}
  \vspace{-1em}
  \Description{User Testing}
\vspace{0.4cm} 
\end{figure*}

\subsection{Task and procedure}
User testing was conducted in a living room with two pairs of participants and one researcher observing the study. The living room was chosen deliberately, due to the assumption that objects in this area are typically on display for others, such as guests or family members, to see~\cite{kirk_opening_2010}.
Before the testing, all participants were asked to take a picture of a significant object in their living room and record a short explanation or story, no longer than one minute, about why the object is important to them. NFC cards featuring images of these objects were then created, and NFC tags were attached to the significant objects at the testing location. 
The study was divided into two phases. In the first phase, participants spent 5 minutes to explore and interact with the device to get a sense of familiarity. The second phase, lasting 15 minutes, focused on participants sharing each other stories through the device, either by scanning the object or inserting a card placed in front of them, see Figure~\ref{fig:Testing}. During this phase, we paid attention to how they interacted with the device and the dynamics of their conversation. After the testing, a semi-structured interview was conducted.

\subsection{Data Collection and Analysis}
During user testing, we observed participant behaviors and their interactions with the device. We documented the interaction using voice recordings, structured field notes, and images. The field notes covered three key areas: 1) device exploration; 2) the story-sharing process; and 3) intriguing or unexpected participant interactions. 
We conducted a semi-structured interview to understand the overall user experience, focusing on investigating whether our design can support human remembering, foster relationship building, and facilitate reflection. We asked participants about the helpfulness of remembering practice after using the device. The voice recordings, field notes, and semi-structured interview responses were analyzed to identify common themes in participants' interactions with the device. We used Fleck's~\cite{fleck_reflecting_2010} framework to analyse the reflection levels observed in the study.

%% file: Body/04-findings.tex
\subsection{Support remembering}
Results showed that the AI-generated questions prompted participants to reflect on their past experiences and mentally revisit these moments. Participant CA-2 expressed,~``\textit{The questions made me think about what the most special moment was in Shizo (the restaurant). You are like trying to bring back those memories, and in my mind, I was highlighting which moments were the most special}''. Those who had difficulty recalling events, such as participant FA-2, found it beneficial. FA-2 mentioned,~``\textit{I am good at forgetting things; this helps me to remind myself and to retrieve the past}''.

Results also showed that the AI-generated questions were perceived as meaningful. For example, our device generated the question,~``\textit{What was the first postcards that you added in your travel journal and what made it so special?}''. Participant CB-1 responded,~``\textit{I think AI generates a good global question, not a yes or no question. I feel like these types of questions are meaningful questions you are supposed to be asking each other rather than normal questions such as how was your day}''. Similarly, when our device asked,~``\textit{What was the most remarkable thing about getting to know your girlfriend during your study period, and how has this influenced your relationship?}''. Participant CA-2 commented,~``\textit{It asks very good questions}''. 
Participants also expressed interest in the next questions generated by AI. One participant (FA-1) mentioned,~``\textit{I really want to know what it is going to ask next}''. Another participant (FRB-1) excitedly said,~``\textit{Are we prepared for the next question?}''.

An even more intriguing finding was that the engagement with Treasurefinder, specifically the questions generated by an LLM, prompted participants to actively retrieve memories. As FRB-1 said,~``\textit{Normally you don't actively think about it, even though it’s visible in the living room. You walk past it, and sometimes you see it, but you don't actively think about it. When questions are asked, you start remembering.}'' Similarly, another participant (FA-1) mentioned,~``\textit{You are consciously retrieving the memories surrounding the everyday life that you normally see in the living room}''.

These findings showed that the interaction with Treasurefinder and the AI-generated questions prompted participants to actively rethink their past, as the questions appeared relevant and meaningful. Interestingly, participants seem to become more conscious of the stories behind their significant objects. Our design facilitated active retrieval of everyday objects that they might easily overlook.

\subsection{Foster new insights about the other person}
Results showed that AI-generated questions in a social setting supported participants to recall their past. These questions also served as conversation starters. For example, participant CB-2 mentioned,~``\textit{Oh, this is a nice way to use it as icebreaker questions}'' and participant CA-1 noted,~``\textit{It definitely initiates a conversation}''. The AI-generated questions encouraged active engagement from participants in the conversation as they continued the discussion and asked follow-up questions; see Table~\ref{tab:conversation}, an example of an ongoing conversation.

Interestingly, the AI-generated questions helped participants understand each other better. As one participant (FB-1) said,~``\textit{I knew you were superstitious, but I didn't know that you highly valued it and that you even have chosen a lucky horse as your significant object. Now that I know the story, I can take that into account for the future}''. Another participant, CB-2, said,~``\textit{I wondered why you have chosen this object. And then you explained it, and it totally made sense. How you associate the object with you}''. It is important to recognize that the participants, spanning friends, couples, and family members, already had strong relationships; therefore, the AI-generated questions did not necessarily enhance these relationships. However, the findings showed that the questions generated by an LLM allowed participants to better understand the other person, providing new insights about the other person.

\begin{table*}
 \vspace{0.5cm} 
  \begin{center}
  \begin{tabular} {p{0.1\textwidth}p{0.8\textwidth}}
    \hline
    \texttt{LLM:} &~``\textit{What was the reason you stopped drawing and painting and how did you find out that these pencils brought back your interest in these activities?}''\\
     \midrule
    \texttt{FA-2:}&~``\textit{Actually two reasons, one is I drew a lot at school, which made it less fun to draw. I didn't feel like drawing in my spare time. And second, that was my wrist injury, which made it impossible to draw...}''\\
    \hline
    \texttt{FA-1:}&~``\textit{But why did you start drawing again?}''\\
    \hline
    \texttt{FA-2:}&~``\textit{Oh yes, thanks to working at Artifact. Shortly, after that I started drawing again because I was surrounded by drawing and painting stuff.}''\\
    \hline
  \end{tabular}
  \vspace{0.4cm} 
  \caption{A transcribed conversation between 2 participants (FA-1 and FA-2)}
  \label{tab:conversation}
   \end{center}
     \vspace{-0.2cm} 
\end{table*}

\subsection{To reflect on the past and current}
Based on Fleck’s 5 (R0-R4) levels of reflection framework~\cite{fleck_reflecting_2010}, we evaluated the participant's reflection levels. Results showed all participants engaged in the first level of reflection, R0, which involved storytelling and recalling events. For instance, CB-2 said, ``\textit{The first postcard that I put in my journal was from 2018 when I went on a trip with my university friends}''.
We also saw that some participants demonstrated a higher level of reflection, R1, where they engaged in reflective thinking rather than simply recalling events. For instance, FRA-2 said, ``\textit{The question made me think and reflect on that moment}'', while FB-1 shared, ``\textit{The device triggered a throwback, which made me think more and reflect on those moments. The questions triggered reflection because they were meaningful questions}''.
Between one pair of participants, we observed a dialogic form of reflection between the past and current situation known as R2. This level of reflection involves a dialogue where individuals share their perspectives and engage in deeper conversational reflection. One participant, FRB-1, reflected on the significance of a simple stuffed animal and how it could bring immense joy compared to the electronic devices commonly used by children today. FRB-2 agreed with this perspective and added insights about their own experience of growing up with a stuffed animal versus the prevalence of digital products. This exchange demonstrates dialogic reflection, where participants engage in a reflective conversation and offer their unique viewpoints.
While we observed that the device could facilitate a form of reflection, we saw that it wasn’t able to facilitate the last two levels of reflection (R3 and R4) by Fleck.
R3 is a reflection with the intent to change one’s behaviors, gain new insights, or reconsider personal assumptions, and R4 relates one’s experience to wider social and ethical implications.

%% file: Body/05-discussion.tex
\subsection{AI-generated questions supports remembering}
The study results demonstrated the good potential that Treasurefinder can support remembering practices, and the AI-generated open-ended questions further enhanced the remembering experience. In particular, AI-generated questions facilitated proactive recollection of past memories. Findings indicate that users actively retrieve memories of objects they easily overlook. However, future iterations might need to consider designing the device into a long-lasting technology to dismiss the effect of another object being overlooked. 

While prior studies have shown that asking open-ended questions supports memory recall~\cite{harris_teaching_2022}, however, many people face challenges in generating open-ended questions due to a lack of interest or knowledge about the topic\cite{harris_teaching_2022,kulkofsky_why_2010}. In our study, participants found that the AI-generated open questions were meaningful and relevant. Therefore, we believe that the  AI's ability to generate relevant questions in different domains allows people to engage more in the reminiscing process. Our findings can inspire researchers, designers, and developers to delve further into the field of using LLMs to prompt individuals to recall their memories as this positively supports remembering.

Nevertheless, there can be risks with AI-generated questions, especially when the story details are incomplete or not elaborate enough. We observed this in test group FA, where the chosen significant object was an ornament symbolizing a miscarried brother. One specific question generated by AI was perceived as emotionless. The AI prompted a question about the miscarried brother, which the participant never got to know about. We observed that AI was unable to generate compassionate questions with the current story. Moreover, the current AI system continuously generates questions based on only one short story stored in the object. This limitation restricts the variation in generating questions. In the future, there is a need to collect more in-depth contextual information to generate a greater variety of AI questions relevant to personal situations, especially when dealing with sensitive information. However, it is crucial to consider data privacy issues and handle personal information carefully and ethically when deploying such systems in the field ~\cite{frauenberger_-action_2017}.

\subsection{Foster new insights about the other person}
Interesting findings showed that testing Treasurefinder in a paired setting enhanced individuals' memory processes. These findings align with prior studies indicating that sharing experiences with others leads to discussions, questions, and reflections~\cite{li_slots-memento_2019,singhal_time-turner_2018}. Most interestingly, findings indicated that the AI-generated questions allowed participants to deepen their understanding of each other, providing new insights about the other person. The findings did not indicate whether it enhanced the relationships among the recruited pairs, given that they already had a strong relationship. However, we see an opportunity to further explore between individuals who may not know each other well, for example, between caregivers and dementia patients or intergenerational relationships. In future research, it would be interesting to explore whether the device could foster relationships among people who are not familiar with each other. 

\subsection{Encourage reflection}
According to Fleck's framework~\cite{fleck_reflecting_2010}, our design supports the foundational reflection (R0-R2) during the remembering process. For example, participants could recall an event without additional reflection (R0), reflect on their experience through reasoning or justifications (R1), and question and interpret different perspectives on past and present situations (R2). However, more research is still needed to investigate whether and how AI-enable reminiscing artifacts could facilitate the last two levels of reflection (R3 and R4). Transformative Reflection (R3) involves seeking fundamental change, aiming to change people's behaviors, gain new insights, or prompt a reconsideration of their personal assumptions. Additionally, there is a need to explore Critical Reflection (R4), which involves critically reflecting on the broader social and ethical implications of experiences, such as moral and ethical issues.

\subsection{Limitations}
We are aware of some limitations of this work. First, the participant's age range was 23-30 years old. Future research could include older adults or people with memory impairments for more valuable insights. Second, the study was conducted in a qualitative setting, which provides rich insights but these context-dependent results may not be generalizable to broader populations.

The device had technological limitations. The quality of the robotic voice output often made it difficult to understand the questions. The voice-recorded stories stored in the NFC tags had to be manually translated into text by the researchers. This process could potentially be automated with a speech-to-text API in the future. Additionally, the compass design was non-functional, making the concept of locating objects inoperative in practice. However, participants did envision this functionality in their living room. Physical objects were attached with NFC tags, which did not preserve the trace of use~\cite{zijlema_preserving_2017}. Future iterations could incorporate a camera in the device with image recognition to detect objects with stories. Most crucially, ethical considerations should be thoroughly addressed when designing such AI-enabled artifacts to support personal reminiscing.

%% file: Body/06-conclusion.tex
This paper presents Treasurefinder, a device integrated with an LLM that generates thought-provoking, open-ended questions with the goal of supporting personal reminiscing. Our study with 12 participants showed the positive effects of using AI-generated questions in the remembering process, including supporting users to recall their past, prompting new insights about each other, as well as facilitating reflection on their past experiences. Moreover, our design facilitated the active retrieval of memories linked to often overlooked but cherished objects. These findings shed light on the possibilities of using LLMs for personal reminiscing, as it supports memory recollection and promotes reflection.